# Proof-of-Concept of Uncompressed 4K Video Transmission from Drone through mmWave

Yoshitaka Takaku†, Yohei Kaieda‡, Tao Yu†, Kei Sakaguchi†
†Department of Electrical and Electronic Engineering, Tokyo Institute of Technology, Japan
‡Intelligent Systems Laboratory, SECOM Co., Ltd, Tokyo, Japan
{takaku, yutao, sakaguchi}@mobile.ee.titech.ac.jp
y-kaieda@secom.co.jp

*Abstract*— Drones are attracting increasing attention in varieties of research fields because of their flexibility and are expected to be applied to a wide range of potential applications, among which the super-high-resolution video surveillance system using drones especially gains the authors' research attention. Surveillance systems using cameras with fixed locations always suffer the blind spots due to the blockage or inappropriate deployments. Instead, by using the drones equipped with cameras, the surveillance performance can be drastically improved due to their high mobilities. The video quality is also a key factor of the surveillance performance. In face recognition, one of the most important surveillance applications, the uncompressed video can greatly improve the detection accuracy, but it is difficult to transmit uncompressed video in real time due to the huge data sizes. To address the issue, we propose to use the ultra-high speed mmWave communication for the video transmission from drones. Moreover, due to the limited battery energy and computing power in drones, we introduce the edge computing and propose to offload all the computation from the drones to the ground station. In addition, a proof-of-concept prototype hardware of the proposed uncompressed 4K video transmission system from drones through mmWave is developed, and the experiments results are consistent with the system design expectations.

*Keywords—5G, mmWave, lens antenna, 4K uncompressed video, face recognition, edge computing, proof-of-concept*

I. INTRODUCTION

In the past decades, unmanned aerial vehicles (UAVs), so-called drones, were mainly adopted in military applications such as unmanned attack, tracking and target localization. And in recent years, thanks to the miniaturization of electronic components and the development of high performance control techniques, drones have become one of the most promising technologies contributing to our daily life, and have been widely employed in varieties of citizen applications, such as sports broadcast, geological survey, target tracking, surveillance, disaster monitoring and logistics[1]-[3], due to their highly flexible and three-dimension spatial mobility. And by 2022, the global market for drone-based business services is expected to be over USD 100 billion per year [4].

Among the wide ranges of drone-based applications, the super-high-resolution video surveillance systems using drone footage especially attract the authors' research attentions. Conventionally, the surveillance systems only employ cameras with fixed locations because of the lacks of mobilities and the limitations of the energy supply and wired communication. As a result, the surveillance systems always suffer the blind spots caused by the blockage or the inappropriate deployment of camera locations.

To address the issue of the limited surveillance coverage, the active and seamless surveillance with no blind spots is expected to be realized by employing camera-equipped drones, which can freely and actively fly in target space. As aerial surveillance system using drone, the case of surveillance system for monitoring the borders and in peace keeping activities is reported[5].

And meanwhile, because the cameras on drones can be deployed in 3D space instead of 2D space, the number of surveillance cameras also increases dramatically for better surveillance performances, which results in huge amounts of videos and images, and it is almost impossible for human security guards to visually check such massive surveillance data in real-time. To solve the problem, research of object and scene recognition based on artificial intelligence (AI) computer vision (CS) have been made to process the surveillance images and videos instead of human[6][7].

Due to the large size of raw video footage, the original video needs to be encoded and then transmitted, especially when the communication is through a wireless link. However, the video quality is also a key factor affecting the performance of a video surveillance system. For example, in the face recognition application, which is one of the most important surveillance applications, the accuracy of face recognition decreases in proportion to the compression rate by applying compression processing to original images[8], and therefore it can greatly improve the detection accuracy by employing the uncompressed ultra-high-resolution video. However, it is impossible to transmit the uncompressed raw video footage in real time through the common bands for drones, e.g., 2.4GHz and 5.7GHz band, because of narrow available bandwidth compared with the huge data size, e.g., 12Gbps of the 4K uncompressed raw footage. The millimeter-wave (mmWave) communication can provide ultra-high-speed and ultra-low-latency wireless data transmission, and has been applied to, e.g., 5G (the 5th Generation Mobile Networks) [9] cellular network since 3GPP Rel.15 and vehicle-to-everything (V2X) networks [10]. To address the above-mentioned bottleneck issue in wireless surveillance video transmission, we propose to transmit the raw video without any compression through 60GHz mmWave communication, which is unlicensed band and has 7GHz continuous available bandwidth.

However, on the other hand, the uncompressed surveillance video footages will inevitably lead to great increase of the computation cost in the surveillance applications such as face recognition. The huge computation amount makes it impossible to perform such processing on the surveillance drones whose energy and computing resource are highly limited. Therefore, in this paper, we also introduce the edge computing [11] and propose to offload all the computation from the drones to the ground station through the mmWave communication.

Moreover, in order to validate the proposed system, a proof-of-concept (PoC) prototype system is developed with the state-of-the-art hardware. The validation experiment is conducted, and the experiments results are consistent with the system design expectations.

This research constructs a mmWave wireless communication system that transmits uncompressed 4K video footage without compression and restoration processing from the surveillance drones in order to improve face recognition accuracy which is performed at the edge side to increase the energy efficiency of the drones. This rest of this paper is organized as follows. Section II presents the system model and explains the general conception of our proposed system. The configuration and structure of the proof-of-concept prototype are presented in detail in Sect. III. Sect. IV presents the validation experiments results. Finally, Sect. V concludes this paper.

## II. SYSTEM DESIGN

This section describes the conception, system design and architecture of the proposed mmWave video transmission system for drones. Furthermore, as one of its surveillance applications, a high-accuracy real-time human recognition system is also proposed.

### A. Overall Architecture of the System

The configuration and architecture of the proposed video transmission system is shown in Fig.1. The whole system is divided into the transmission side (drone) and receiver side (ground station) using edge computing. In the transmission side, surveillance drones equipped with 4K cameras monitor the target area and objects (e.g., stadium and human). And the 4K uncompressed video is transmitted from the drone to the ground station through mmWave communication.

To illustrate the advantages, we applied it to a real-time face recognition system. In the ground station, human detection is performed by analyzing the received 4K uncompressed video in real-time. Thanks to ultra-fast communication and high video resolution, the accuracy of face recognition can be greatly increased. Moreover, because the edge computing is introduced and all computation is offloaded from the drone to the ground station, the low latency face detection and high energy stability can be archived in this architecture.

The typical use-case of this system is expected to be the patrol for the mega event, e.g., the stadium as shown in Fig.2. The drone will fly freely around the space and monitor crimes and terrorism around the stadium. The drone can fly around freely to eliminate blind spots and monitor a wider field of view. In addition, various advantages can be considered to detect human by AI. First, it is possible to perform more precise patrol than monitoring with human eyes directly. Second, In this research, we will discuss the effectiveness of face recognition, but in the future it will be possible to monitor how people behave by AI and prevent crimes, or to detect crimes and follow up by drones[12][13].

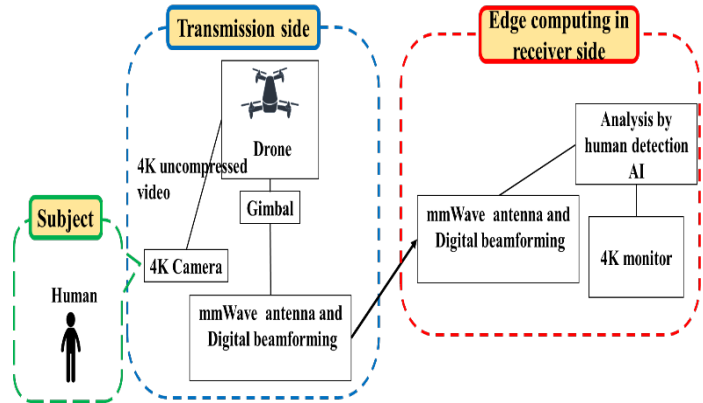

Fig. 1. System architecture.

### B. 4K Uncompressed Video Transmission from Surveillance Drones Using mmWave

In common video surveillance systems, the video compression such as H.264 and H.265 is conducted due to the large size of the raw video footage. For example, when video of 4K/60fps is transmitted without compression, it yields huge data size and requires about a data rate of 12Gbps. In comparison, it is possible to reduce bit rates by compressing the video. However, uncompressed method has better points than compression method at the expense of having huge data sizes. First, as mentioned above, the accuracy of face recognition decreases in proportion to compression rate. Second, compression method needs to perform computation processing, which results into higher computation latency. Finally, power consumption also increases by such computation processing. For these reasons, we select uncompressed 4K video for the system's improvement of especially object recognition rate, latency and power consumption.

In order to transmit uncompressed video, high-capacity communication link is required. Conventionally, frequency that is used for drone video transmission is 2.4GHz or 5.7GHz band but it is difficult to ensure a sufficient bandwidth to transmit uncompressed video. This problem can be solved by using 60GHz band that guarantees a large bandwidth of 2.16GHz[14], and thus uncompressed video transmission is made possible. mmWave has large propagation attenuation. Therefore, directional antenna of converged beam-width for long-distance communication is required. Basically, there are two types of beamforming. One is digital beamforming that switches the beam digitally, the other is mechanical beamforming that switches the beam in analog manner. If ground access point (AP) communicates with drone at a longer distance, the angle to move the antenna's beam instantaneously becomes smaller for high-speed movement of drone. In this paper, targeting short distance communication, the faster drone moves, the greater the angle at which antenna's beam should be re-directed instantaneously as depicted in Fig.3. For that reason, we select digital beamforming against analog beamforming owing to its higher capability of angular adaptability. Also, we suppose the free-space propagation model without multipath at short distance.

*C. Human Detection by Edge computing*

A case where AI processing at the ground station is performed at the transmission side are also possible but it is supposed that power consumption significantly increases by AI processing and drone payloads increase. Therefore, by performing AI processing at the ground station, computation is performed without limitation of power consumption, so AI with various analysis capabilities can be installed, such as tracking, not just face recognition. In addition, processing in edge computing leads to low latency and early response to abnormal situations.

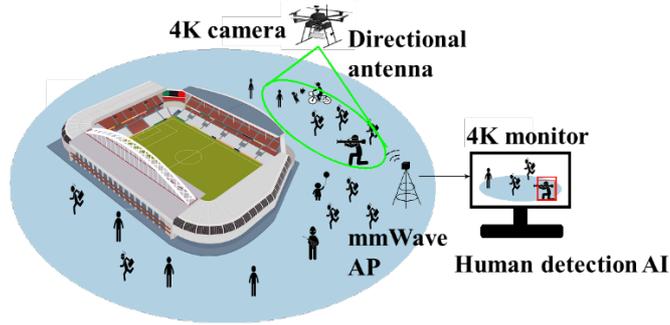

Fig. 2. Use-case by drone-based surveillance.

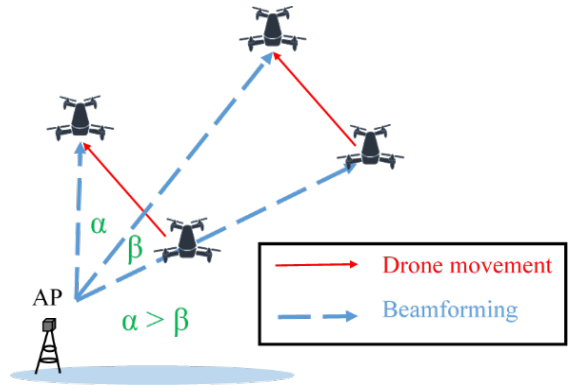

Fig. 3. Angle at which antenna is directed against drone movement.

### III. PROOF-OF-CONCEPT PROTOTYPE

In order to evaluate the whole system, experiments of 4K uncompressed video transmission and human detection was conducted. 4K uncompressed video transmission experiment was conducted by flying drone. However, human detection was conducted on the desk without flying the drone because it was difficult to perform both experiments at the same time in this experiment hardware specifications and environment. This section describes configuration and structure of the proof-of-concept prototype in each experiment.

*A. Hardware Prototype of Uncompressed 4K Video Transmission System*

*1) mmWave communication in drones*

In this research, lens antenna shown in Fig.4 is used as a directional antenna. It has been difficult to mount on drones due to size and weight issues but lens antenna used in this research is about 1kg as a whole and is small and lightweight, and thus can be equipped on drones.

Table I shows antenna parameter with and without lens. Scan angle refers to how much beam can be scanned horizontally and vertically. This antenna realizes high directivity gain, can be used for wide range of communication by combining with digital beamforming.

Numerical analysis is conducted to show how much the transmission distance is improved with and without the lens. Table II shows each simulation parameter in this paper. Parameters are referred to IEEE802.11ad standard [14]. the Error Vector

Magnitude (EVM) value of Modulation and Coding Scheme (MCS) in [14] is used as an evaluation criterion. In addition, the inverse of EVM can be interpreted as SNR (Signal-to-noise-ratio). This lens antenna can transmit MCS 12 at maximum.

Simulation results are summarized in Fig.5. When SNR is 21dB, transmission distance of antenna gain 17.5dBi is 21.6m and antenna gain 25.4dBi is 132.6m. Therefore, it can be seen that transmission distance has increased significantly.

*2) Drone and Camera, Gimbal*

Figure 6 shows a photo of drone with mmWave antenna and gimbal used in this research, and Table III shows the drone and camera's specifications. A gimbal is usually used on the camera to take clear images or vides. In this research, the mmWave antenna is attached to a mechanical rotatable gimbal to compensate the shifting of polarization plane caused by, e.g., drone shaking and wind, especially when the mmWave antenna is linear polarization. Therefore, gimbal moves mmWave antenna three-dimensionally and correct the shift of polarization plane by the amount of shift. Image resolution used in this experiment is 3840×2160 called Ultra high definition (UHD). Frame rate is 7.5fps and bit rate is about 1.5Gbps. The reasons why the frame rate is lower are that the maximum throughput of the lens antenna is 1.5Gbps, and video transmission is not possible when a higher frame rate is set.

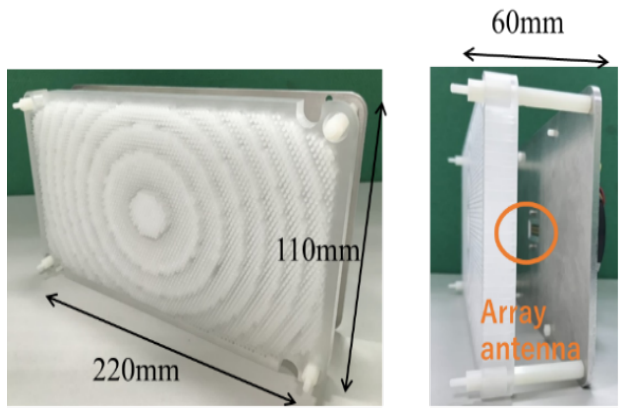

Fig. 4. Lens antenna.

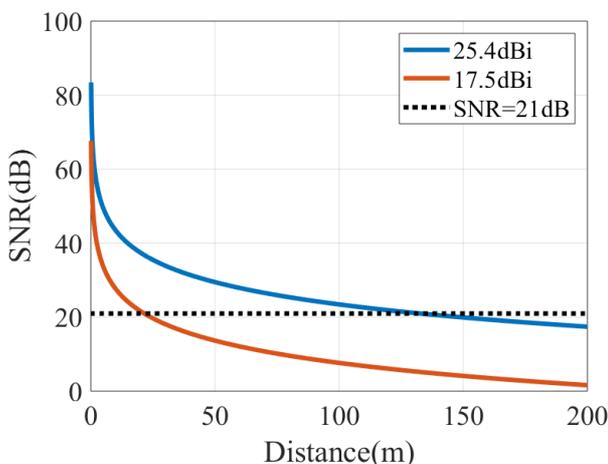

Fig. 5. Transmission distance.

TABLE I. LENS ANTENNA PARAMETERS.

| Antenna | Gain | Tx power | Scan angle |
|---|---|---|---|
| Lens antenna | 25.4dBi | 10dBm | +/- 13.5 deg (Hor.), +/- 7deg (Ver.) |
| Array antenna | 17.5dBi | | +/- 49 deg (Hor.), +/- 19.5 deg (Ver.) |

TABLE II. SIMULATION PARAMETERS.

| Parameter | Value |
|---|---|
| Carrier freq. | 60GHz |

| Bandwidth | 2.16GHz |
|---|---|
| Link | Up |
| Propagation loss | $20\log_{10}(4\pi d/\lambda)$ |
| Noise power density | -174dBm/Hz |
| Noise factor | 10dB |

TABLE III. DRONE AND CAMERA SPECIFICATIONS.

| Device Name | Relevant Parameters |
|---|---|
| | *Specifications* |
| Drone ACSL-PF1 | Size : 117cm×106cm×41.8cm, Weight : 6.4kg, Flight time : 45min, Payload : 3kg |
| Camera See3CAM_130 | Size(L×W×H) : 80mm×15mm×9.67mm, Image resolution : 4224×3156 and 3840×2160, 1920×1080 |

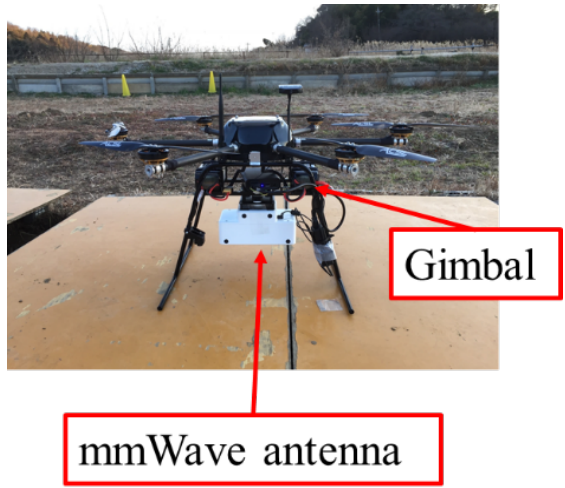

Fig. 6. mmWave lens antenna drone.

*B. Human Detection Algorithm and Software*

In human detection experiment, Open Source Computer Vision Library (OpenCV) is used as image processing software. Haar-like features and cascade classifiers are used as human detection algorithms. Haar-like feature is calculated from the difference in brightness and darkness of the image[15][16]. The cascade classifier extracts feature quantities during learning and generates a classifier[15]. In addition, the detector "haarcascade_frontalface_default.xml" file that has been learned in advance on OpenCV is used as a human detector in this experiment[17].

IV. EXPERIMENT RESULTS

In this section, we describes experiment configuration and results of 4K uncompressed video transmission using drone with mmWave antenna and human detection using AI.

*A. Video Transmission Experiment Results*

In the experiment, a drone was made to hover right above the ground AP as depicted in Fig.7. As a result, we succeeded in transmitting real-time 4K uncompressed video captured by a 4K camera mounted on drone to a ground AP up to 100m[18][19]. It was confirmed that high-quality 4K uncompressed video transmission was possible. Bit rate is about 1.5Gbps. Aerial photo from height 100m is shown in Fig.8.

*B. Human Detection Experiment Results*

Fig.9(a) shows the configuration of human detection experiment. The faces are captured by a 4K or 2K camera, and images are transmitted using millimeter waves. After that, face recognition is performed by human detection AI on the receiving side, and the result is output to 4K Monitor.

Fig.9(b) shows experiment hardware configuration. This experiment was conducted in Fig.9(a)'s configuration. The camera used in the drone with mmWave antenna 4K uncompressed video transmission experiment can be switched between uncompressed 2K and 4K. As faces to be recognized, 136 faces of various sizes obtained from google search are recognized at once (Face part in the Fig.9(b)). In addition, in order to reproduce a scene that recognizes distant faces, the experiment is conducted by reducing the target face.

In this experiment, a plurality of faces captured by a 2K or 4K camera are transmitted by a millimeter-wave antenna, and the faces are judged using the cascade classifier on the receiving side.

Table IV summarizes the experiment results that show the number of recognized faces in uncompressed 2K and 4K respectively[20]. From the results of the average number of the recognized faces, it can be seen that uncompressed 4K can greatly improve the number of face recognition compared to uncompressed 2K. In addition, farther objects can be recognized by increasing the resolution, and a wider range of objects can be captured. By using uncompressed 4K to increase the resolution, it is possible to recognize distant objects, which is suitable for this system.

*C. Power Consumption by human detection*

We measured power consumption of hardware in this experiment configuration. Power consumption at receiver side increases to 17.3W from 3.8W when 4K uncompressed video transmission and human detection AI processing are performed from the state where the personal computer (PC) is only turned on. In order to make human detection with higher accuracy and real-time, it is necessary to improve PC specifications. In this experiment, the number of core in PC is 4 and clock frequency is 1.6GHz. In general, if the number of core in central processing unit (CPU) is increased, power consumption increases. To solve this problem, lowering clock frequency usually lead to reduce power consumption[21]. Also, graphic processing unit (GPU) is often used for image processing but GPU consumes more power than CPU[22]. As a result, it is expected that further power consumption will be consumed as hardware specifications are increased, and it is difficult to perform AI processing drone side with flight time of less than an hour.

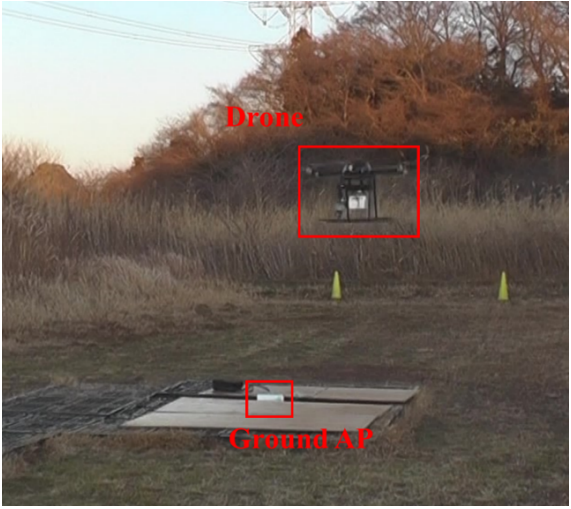

Fig. 7. 4K uncompressed video transmission experiment configuration.

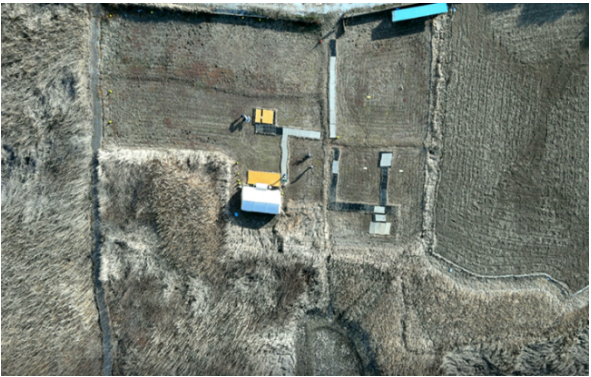

Fig. 8. The photo from height 100m.

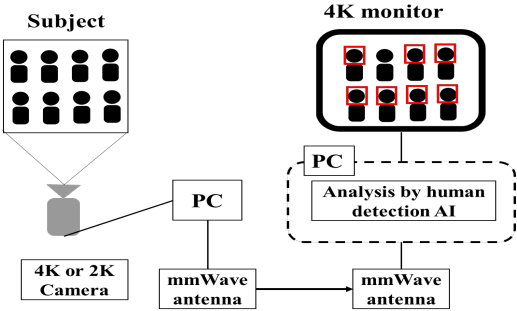

(a) Human detection experiment configuration.

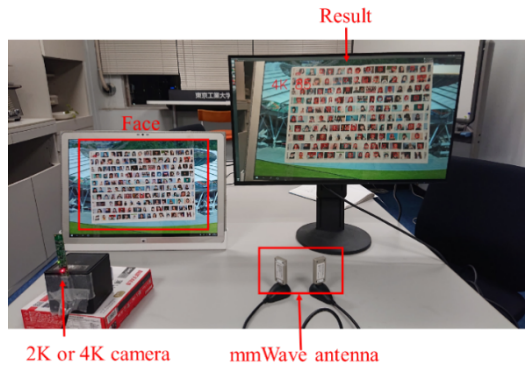

(b) Human detection experiment hardware configuration.

Fig. 9. Human detection experiment.

TABLE IV. NUMBER OF FACE DECTION

| Resolution | Number of faces | Average number of the recognized faces |
|---|---|---|
| Uncompressed 2K | 136 | 45.5 |
| Uncompressed 4K | 136 | 86.2 |

## V. CONCLUSION

In this research, we proposed 4K uncompressed video transmission system using mmWave drone, and conducted 4K uncompressed video transmission experiment and human detection experiment with different resolutions. As a result, we successfully transmitted 4K uncompressed video from 100m above the sky. Furthermore, it was confirmed that the detection accuracy was greatly improved compared to 2K even if the person captured by the camera became smaller by separating the distance. Also, AI computation results in increase of power consumption. This result shows the effectiveness of AI processing on the receiving side, not on the drone side.

Our future work is increasing frame rates of 4K uncompressed video and transmit smoother video. In addition, currently, we have confirmed the difference in recognition rate due to the difference in resolution, but we will also check the difference in recognition rate due to compression and uncompressed video.

ACKNOWLEDGMENT

This research leading to these results is jointly funded by the European Commission (EC) H2020, the Ministry of Internal affairs and Communications (MIC) in Japan under grant agreements N° 723171 5G-MiEdge in EC and 0159-0077, 0155-0074 (MiEdge+) in MIC. In addition, This research leading to these results is funded by the Ministry of Internal affairs and Communications (MIC) in Japan.

REFERENCES

[1] J. Gu, T. Su, Q. Wang, X. Du and M. Guizani, "Multiple Moving Targets Surveillance Based on a Cooperative Network for Multi-UAV," in IEEE Communications Magazine, vol. 56, no. 4, pp. 82-89, April 2018.

[2] "Amazon to deliver purchased by drone "within months"" https://www.dezeen.com/2019/06/06/amazon-prime-air-drone-news/

[3] R. Ma, X. Li, M. Sun and Z. Kuang, "Experiment of Meteorological Disaster Monitoring on Unmanned Aerial Vehicle," 2018 7th International Conference on Agro-geoinformatics (Agro-geoinformatics), Hangzhou, 2018, pp. 1-6.

[4] "Drones", https://www.goldmansachs.com/insights/technology-driving-innovation/drones/, [Online]

[5] Z. Zaheer, A. Usmani, E. Khan and M. A. Qadeer, "Aerial surveillance system using UAV," 2016 Thirteenth International Conference on Wireless and Optical Communications Networks (WOCN), Hyderabad, 2016, pp. 1-7.

[6] H. Chen, C. Liu and W. Tsai, "Data Augmentation for Cnn-Based People Detection in Aerial Images," 2018 IEEE International Conference on Multimedia & Expo Workshops (ICMEW), San Diego, CA, 2018, pp. 1-6.

[7] A. Rohan, M. Rabah and S. Kim, "Convolutional Neural Network-Based Real-Time Object Detection and Tracking for Parrot AR Drone 2," in IEEE Access, vol. 7, pp. 69575-69584, 2019.

[8] Article in Proc. of the 79th National Convention of IPSJ (in Japansese) Available at: http://id.nii.ac.jp/1001/00180807

[9] V. Frascolla, K. Yunoki, S. Barbarossa, F. Miatton, G. K. Tran, et al., "5G-MiEdge: Design, Standardization and Deployment of 5G Phase II Technologies," Proc. of IEEE CSCN 2017, Helsinki, Finland, Sep. 2017.

[10] Z.Li, T.Yu, R.Fukatsu,G.K.Tran,K.sakaguchi, "Proof-of-Concept of a SDN Based mmWave V2X Network for Safe Automated Driving," in IEEE Globcom2019.

[11] H. Nishiuchi, G. K. Tran, K. Sakaguchi, "Performance Evaluation of 5G mmWave Edge Cloud with Prefetching Algorithm," IEEE VTC2018-Spring, Jun. 2018.

[12] S.Xie, X.Zhang,J.Cai, "Video crowd detection and abnormal behavior model detection based on machine learning method," Neural Computing and Applications (2019) 31 (Suppl 1):S175–S184

[13] Y.Imamura, S.Okamoto, J.H.Lee, "Human Tracking by a Multi-rotor Drone Using HOG Features and Linear SVM on Images Captured by a Monocular Camera," Proceedings of the International MultiConference of Engineers and Computer Scientists 2016 Vol I, IMECS 2016, March 16 - 18, 2016, Hong Kong

[14] IEEE 802.11 ad, "Part 11: Wireless LAN medium access control (MAC) and physical layer (PHY) specifications," pp.2436-2496, Dec. 2016.

[15] P. Viola and M. Jones, "Rapid object detection using a boosted cascade of simple features," Proceedings of the 2001 IEEE Computer Society Conference on Computer Vision and Pattern Recognition. CVPR 2001, Kauai, HI, USA, 2001, pp. I-I.

[16] R. Lienhart and J. Maydt, "An extended set of Haar-like features for rapid object detection," Proceedings. International Conference on Image Processing, Rochester, NY, USA, 2002, pp. I-I.

[17] "haarcascade_frontalface_default.xml",https://github.com/opencv/opencv/blob/master/data/haarcascades/haarcascade_frontalface_default.xml, [Online]

[18] "Drone transmits uncompressed 4K video in real time using millimeter wave tech", https://www.titech.ac.jp/english/news/2019/044615.html, [Online]

[19] "[Demo] Transmission Uncompressed 4K Video from Drone through Millimeter-Wave Communication",https://www.sakaguchi-lab.net/2019/04/20/demo-transmission-uncompressed-4k-video-from-drone-through-millimeter-wave-communication/, [Online]

[20] "[Demo] Face recognition by 4k video through mmWave communication", https://www.sakaguchi-lab.net/2019/09/21/demo-face-recognition-by-4k-video-through-mmwave-communication/, [Online]

[21] H. Kataoka, D. Duolikun, T. Enokido and M. Takizawa, "Power Consumption and Computation Models of a Server with a Multi-core CPU and Experiments," 2015 IEEE 29th International Conference on Advanced Information Networking and Applications Workshops, Gwangiu, 2015, pp. 217-222.

[22] T. Matsumoto, S. Yamaguchi and T. Sakai, "A Study on Improving Power-Consumption Performance Ratio in GPGPU Computing," 2011 Second International Conference on Networking and Computing, Osaka, 2011, pp. 288-290.